\documentclass{emulateapj}
\usepackage{psfig}
\usepackage{graphicx}
\usepackage{natbib}
\submitted{Submitted to The Astrophysical Journal, Letters (March 15,
2004)}

\advance \voffset by 0.00cm\relax

\shorttitle{First Stellar Binary Black Holes: Strongest Gravitational Wave
Burst Sources}
\shortauthors{Belczynski, Bulik \& Rudak}
\begin{document}

\title{First Stellar Binary Black Holes: Strongest Gravitational Wave Burst Sources}

\author{Krzysztof Belczynski\altaffilmark{1,2}, Tomasz
Bulik\altaffilmark{3}, Bronislaw Rudak\altaffilmark{4}}

\affil{
     $^{1}$ Northwestern University, Dept. of Physics \& Astronomy,
            2145 Sheridan Rd., Evanston, IL 60208\\
     $^{2}$ Lindheimer Postdoctoral Fellow\\
     $^{3}$ Nicolaus Copernicus Astronomical Center,
            Bartycka 18, 00-716 Warszawa, Poland;\\
     $^{4}$ Nicolaus Copernicus Astronomical Center,
            Rabianska 8, 87-100 Torun, Poland;\\
      belczynski@northwestern.edu, bulik@camk.edu.pl, bronek@ncac.torun.pl}

\begin{abstract}
Evolution of first population of massive metal-free binary stars is followed.
Due to the low metallicity, the stars are allowed to form with large
initial masses and to evolve without significant mass loss.
Evolution at zero metallicity, therefore, may lead to the formation of
massive remnants. In particular, black holes of intermediate-mass ($\sim
100-500$\ M$_\odot$) are expected to have formed in early Universe, in contrast
to the much lower mass stellar black holes ($\sim 10$\ M$_\odot$) being
formed at present.
Following a natural assumption, that some of these  Population III stars
have formed in binaries, the physical properties of first stellar binary
black holes are presented. We find that a significant fraction of
such binary black holes coalesces within the Hubble time.
We point out that burst of gravitational waves from the final coalescences
and the following ringdown of these binary black hole mergers can be 
observed in the interferometric detectors. 
We estimate that advanced LIGO detection rate of such mergers is at least  
several events per year with high signal to noise ratio ($\gtrsim 10$). 
\end{abstract}


\section{INTRODUCTION}
The properties of Population III stars have stirred
a lot of interest in the recent years.
It has been realized that zero metallicity stars
with masses up to several hundred of solar masses
are stable \citep{2001ApJ...550..890B}.
\citet{2002ApJ...567..532H} estimated black hole (BH) 
masses formed in the evolution of high-mass metal-free 
stars.
For initial masses above $40$\ M$_\odot$
the remnant is a BH with the mass essentially
the same as the progenitor with the exception
of stars with initial masses between $140$ and $260$\ M$_\odot$
which undergo a pair instability supernova (SN) explosions and 
leave no remnant at all.
\citet{2003ApJ...591..288H} found that these conclusions hold
also for low, non-zero metallicity stars.
Numerical studies of collapse and fragmentation
of a metal free gas in the early Universe  
\citep{1999ApJ...527L...5B,2002ApJ...564...23B,2003ApJ...589..677O}   indicate
 that the initial mass function (IMF) of Population III stars
is top heavy, and might be bimodal with the high mass
peak around  $100$\ M$_\odot$
\citep{1998MNRAS.301..569L,2001ApJ...548...19N,2003PASP..115..763C}.
Every known stellar system contains a considerable fraction of binaries. 
Therefore it seems quite natural to allow for a possibility that some
fraction of massive Population III stars formed in binaries as well, 
and then investigate their subsequent evolution.
There is a possibility that these systems form massive black hole black 
hole (BH-BH) binaries which in turn may be observable in gravitational waves 
by the interferometric detectors.

In this paper we consider the properties of Population III binaries
with the initial component masses in the range $100-500$\ M$_\odot$.
In \S\,2 we describe the model of evolution of metal-free systems leading 
to formation of massive BH-BH binaries.  
In \S\,3 we present characteristic properties of the BH-BH population 
and discuss the observability of their mergers. Conclusions and summary of results
are given in \S\,4.

\section{Evolutionary Model}
A simple model of the evolution of Population III 
stars was constructed. Lacking the observational input, we used a set 
of recent calculations for metal-free single stars and then combined 
them with the basic binary evolutionary prescriptions.

Using the numerical calculations of stellar tracks for Populations III stars
\citep{2001ApJ...550..890B, 2001AnA...371..152M, 2002AnA...382...28S} 
we obtained the evolutionary timescale and radial expansion history of 
a star as a function of its initial mass.
After the core hydrogen exhaustion, we calculate He-core core mass for a
given star using the approximate empirical formula of \citet{2002ApJ...567..532H}.
Once a star has finished its nuclear evolution, we follow the work of 
\citet{2003ApJ...591..288H} to decide on core collapse outcome. Depending on
the initial star mass, the low-metallicity massive star may either collapse
directly and form a BH (without accompanying SN explosion) or be
entirely disrupted (no remnant left) in a pair instability SN (stars within
initial mass range of 140--260 M$_\odot$).
In the former case, the total mass of collapsing star forms a BH, and we
assume that there is no natal kick associated with the direct BH formation. 
The pulsational pair instability is also taken into account as it may remove 
part of the star envelope just prior to the collapse
\citep{2003ApJ...591..288H}. We assume that half of the envelope is lost for
the stars with  initial mass in the  range: 100-140 M$_\odot$.

The orbit of each binary is assumed to be circular, and the orbital
separation is drawn from a distribution flat in logarithm with the maximum value
of $10^6$\ R$_\odot$ \citep{1983ARA&A..21..343A}.
Since little is known about the shape of the IMF of 
Population III stars we adopt a power law shape with $\alpha=-2$ 
exponent and draw the primary (initially more massive component) mass 
from such a distribution in the range of $100-500$\
M$_\odot$. The secondary mass is obtained through mass
ratio (secondary/primary) which is drawn from a flat distribution.
Radial expansion of the components is followed, and in the case of Roche lobe
overflow (RLOF) we apply one of the following prescriptions to calculate the
outcome. For unevolved (main sequence) donors we assume that the RLOF will
always lead to the component merger, thus terminating further evolution and
aborting potential BH-BH formation.
For evolved donors, we check whether the RLOF may lead to dynamical instability.
If the donor mass is  larger than twice the accretor mass ($q_{crit}=2$), we 
apply standard common envelope 
prescription \citep{1984ApJ...277..355W} with 100\% efficiency of inspiral 
orbital energy conversion into the envelope ejection. Otherwise, we use the
non-conservative evolution with half of the transferred material leaving the
system with the angular momentum specific to a given binary 
\citep{2002ApJ...572..407B}.
In the case of accretion, the unevolved stars are rejuvenated and may reach
larger radii than their initial mass would have suggested, while the evolved
stars are allowed only to increase their mass.
Once a BH-BH binary is formed, the orbit decay time due to a gravitational
wave emission is calculated \citep{1964PhRv..136.1224P}.

\section{Results}

\subsection{Formation and Properties of BH-BH Population}
A large set ($N_{\rm tot}=10^6$) of Population III binaries, described in
the previous section, is evolved.
Evolution leads to efficient formation of BH-BH systems (38\%), despite 
the fact that significant fraction of binaries cease to exist when one
of the components is disrupted in a pair instability SN (46\%), or
components merge in RLOF event (16\%).

\begin{deluxetable}{clr}
\tablewidth{230pt}
\tablecaption{BH-BH FORMATION CHANNELS}
\tablehead{Channel & Evolutionary Sequence\tablenotemark{a} & Efficiency}
\startdata
bhbh01& BH1 BH2         & 0.67\\
bhbh02& MT1 BH1 CE2 BH2 & 0.15\\
bhbh03& MT1 BH1 MT2 BH2 & 0.07\\
bhbh04& BH1 MT2 BH2     & 0.05\\
bhbh05& all others      & 0.06\\[-0.3cm]
\enddata
\label{channels}
\tablenotetext{a}{BH1/BH2: first/second BH formation. CE: common
envelope, MT: non-conservative RLOF, where 1 or 2 denotes the donor,
either primary or secondary, respectively.}
\end{deluxetable}

Major evolutionary channels along with BH-BH formation efficiencies
are listed in Table~1.
Most of the primordial systems form on wide orbits and never interact
(channel bhbh01).  Systems formed on the tight orbits interact twice, 
as first the primary then the secondary overfill their Roche lobes (bhbh02, 
bhbh03). Systems formed on the intermediate orbits interact only once; 
depending on the maximum component radii and the mass ratio RLOF is
initiated either by the secondary (bhbh04) or the primary (within bhbh05).
 
\begin{figure}
\begin{center}
\psfig{figure=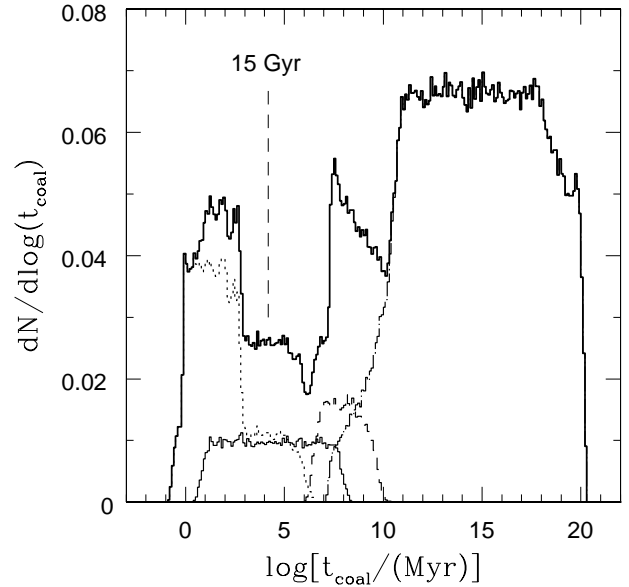,width=3.5in}
\caption{
The distribution of BH-BH coalescence times (thick solid line) 
normalized to unity.
Additionally, main subpopulations, which have evolved to BH-BH stage 
along different evolutionary channels, are presented with thin lines; 
bhbh01 long dashed dotted, bhbh02 dotted, bhbh03 solid, bhbh04 
short dashed. Note that the subpopulation of systems formed along 
bhbh05 channels with $\log(t_{\rm coal}) \sim 3-10$ is not shown. 
For definition of channels see Table 1.
}
\label{pl1}
\end{center}
\vspace{-0.7cm}
\end{figure}

In Figure~1 we present distribution of the coalescence times for the BH-BH
populations. A  significant fraction of BH-BH systems (0.17)  coalesces
within the Hubble time (15 Gyr).
Binaries with short and intermediate initial periods interact and
tend to either form tight BH-BH or merge in RLOF. Various subpopulations
of BH-BH systems are shown in Fig~1. The stronger (CE in case
of bhbh02 as compared to non-conservative mass transfer) or more 
frequent interactions (two interactions in bhbh03 versus one in
bhbh04) lead to shorter coalescence time of BH-BH binary.
Large coalescence times are found for systems which never interacted,
and basically remained on the unchanged wide orbits throughout the
evolution (bhbh01). Systems with large coalescence times dominate the
population, since they have formed without interactions (wide orbits)
and avoided possibility of component merger in RLOF events.

Binary BHs found in our calculation cover a wide range of masses;
50--500 and 40--620 M$_\odot$ for the first and the second BH formed in a
system, respectively.
Total mass of BH-BH binaries spans the wide range $M=100-1000$\ 
M$_\odot$, with most of the systems forming with $M=100-200$\ M$_\odot$. 
The distribution of masses of the remaining  population
peaks  at $M=350$\ M$_\odot$ and has a tail extending  up to $M=1000$\ M$_\odot$.

\begin{figure}
\includegraphics[width=0.95\columnwidth]{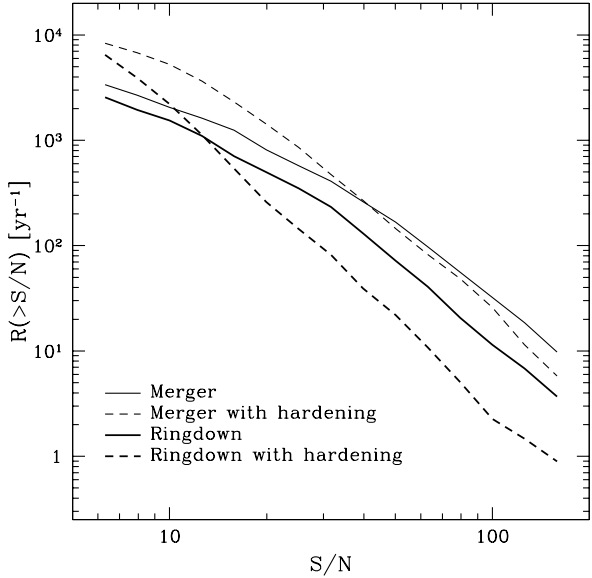}
\caption{The expected observed coalescence rate of Population III 
BH-BH binaries as a function of signal to noise ratio for advanced 
LIGO detector.
The thick lines correspond to detection of the ringdown signal
and the thin lines to detection of the merger (burst) signal.
Solid lines are for isolated orbital decay of BH-BH systems, while 
dashed lines represent the case when binaries are hardened in dense
environment (see text for details).
}
\label{rsn}
\vspace{-0.4cm}
\end{figure}

\subsection{Observability in Gravitational Waves}

A coalescence of two black holes proceeds through three consecutive 
phases: inspiral, merger, and ringdown.
The signal to noise ratio ($S/N$) from each phase of a coalescence in 
the existing and near-future interferometers has been calculated by
\citep{1998PhRvD..57.4535F}.
The typical total mass of a black hole binary in the population
considered in this {\em Letter} is quite large and reaches above a 
few hundred M$_\odot$.
The inspiral signals in interferometric detectors for such massive binaries 
is very low since it falls outside the maximum sensitivity range. 
However, the merger and ringdown signals for the advanced LIGO peak 
for the redshifted total masses between 100 and 2000 M$_\odot$.

We calculate the observed BH-BH coalescence rates following the formalism 
in \citet{2004A&A...415..407B}, see also \citet{2003ApJ...589L..37B}.
We estimate the comoving rate of Population III star formation 
$R_{\rm sfr}$ which would satisfy the following assumptions: 
{\em i)} formation of Population III stars occurs at a constant rate 
at redshifts $z$ from $30$ to $10$, {\em ii)} $f_{\rm mass} =
10^{-3}$ of
baryonic matter goes into the Population III stars \citep{2001ApJ...551L..27M}.
We adopt a flat cosmology model with density parameter of matter
$\Omega_m=0.3$, density parameter of cosmological constant
$\Omega_\Lambda=0.7$ and for Hubble constant $H_0=100 h \, {\rm km}
\ {\rm s}^{-1} {\rm Mpc}^{-1}$, with $h = 0.65$. We obtain
$R_{\rm sfr} \simeq 1.4 \times 10^{-2} {{\rm M}_\odot} 
{\rm Mpc}^{-3} {\rm yr}^{-1}.$
Furthermore, we assume that the binary fraction in Population III
stars is $f_b=0.1$, and that the IMF below $100\,{\rm M}_\odot$ 
is flat, and extends down to $1{\rm M}_\odot$.
We first calculate the differential merger rate as a function of redshift
$d f_{coal}(z) / d{M }$ taking into account
the delay between formation and coalescence  due to gravitational inspiral.
The differential coalescence rate per unit observed mass is
\begin{equation}
{dR \over d{ M_{obs}}} = \int_0^{z_{ M} }
{d f_{coal}(z) \over d{M} } {1\over 1+z} {dV\over dz}
dz
\label{obsrate}
\end{equation}
where $M_{obs}=M(1+z)$ is the observed (redshifted) total mass,
$z_{M}$ is the maximum redshift out to which a binary is observable,
and ${dV/dz}$ is the comoving volume element. The maximum redshift 
$z_{M}$ is estimated using the $S/N$ values estimated by
\citet{1998PhRvD..57.4535F} for the advanced LIGO detector for merger 
and ringdown phases.

We present the expected rates for advanced LIGO as a function of the value of 
the $S/N$ threshold as solid lines in Figure~\ref{rsn}. The curves calculated 
assuming  detection in merger and ringdown phases are similar because of
similarity of the dependence of $S/N$ one the masses of the system in the two 
cases. However, it has to be noted that the merger $S/N$ is rather uncertain 
since its calculation requires precise prediction of detector noise curve as
well as knowledge of highly uncertain binary parameters (e.g., BH spins). 
Thus the detection efficiency calculated with the use of the predicted 
merger $S/N$ may be much smaller than presented in Figure~\ref{rsn}.
On the other hand the ringdown signal is much better constrained and
easier to predict. 

Population III BH-BH binaries might have populated dense stellar 
environments in young galaxies. If this was the case their coalescence  
times could have been significantly shortened due to the additional 
orbital decay by three-body interactions. We estimate the observed 
coalescence rates in the alternative model in which we allow for 
additional orbital decay due to the dynamical hardening through simple 
orbital shrinkage for all BH-BH binaries. Orbits are tightened by 
factor of 10, causing the decrease of the coalescence times by factors 
$\sim 10^4$. The corresponding rates are shown in Figure~\ref{rsn}. 
Hardening does not have a strong impact on the observed rates because 
of two opposing effects. On one hand the tight binaries merge at higher 
redshifts and their detectability drops. On the other, the large 
population of wide non coalescing (for standard model, see Fig.~\ref{pl1}) 
binaries is shifted to shorter coalescence times, and in particular 
some may add to the predicted detection rate when hardening is included.

\begin{figure}
\includegraphics[width=0.95\columnwidth]{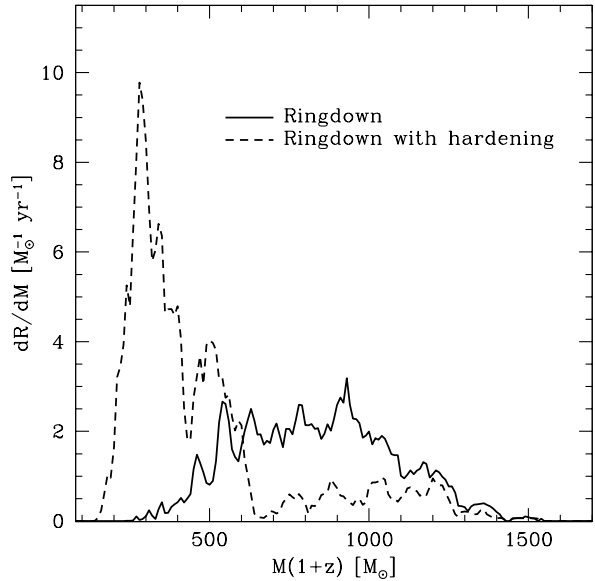}
\caption{The differential rate as a function of observed (redshifted)
total mass of BH-BH system. In this calculation we required a detection
with high signal to noise ratio ($S/N>10$).}
\label{rate}
\vspace{-0.4cm}
\end{figure}

Redshifted total mass of the system is the principal quantity that can 
be inferred from the observation of a ringdown signals. 
In Figure~\ref{rate} we present the observed rate as a function
of the redshifted total mass, requiring a detection with $S/N=10$. 
The typical redshifted total mass lies between $600$ and   $1000\,{\rm M}_\odot$.
In the case of the alternative model with hardening included the 
typical value shifts down to $\approx 300\,{\rm M}_\odot$. This is 
due to the fact that originally long lived binaries contain lighter
black holes. A calculation using the merger signals leads to  very
similar results.

\section{Discussion}
The principal result of the calculations is presented in Figure~\ref{rsn}.
The expected coalescence rate of Population III intermediate-mass BH-BH
binaries is high. We predict that advanced LIGO should observe above a 
thousand strong ($S/N \gtrsim 10$) events per year. What are the
uncertainties of this prediction?

In our calculation we have assumed that Population III stars were formed
at redshifts $z$ from $30$ to $10$ out of $f_{\rm mass} = 10^{-3}$ of the baryon mass
contained in the Universe and that the binary fraction of the initial population
was $f_b=0.1$. The exact length and duration of the Population III star
formation era does not affect the rate calculation.
However, the rate scales linearly with $f_{\rm mass}$ and $f_b$, and although
we have chosen rather conservative values for these two quantities, the rate
may decrease if the adopted values were significantly lowered.

The IMF of Population III stars is another unknown. Numerical investigations
show that the IMF leans towards massive stars. We assumed rather steep IMF
(with the slope $\alpha=-2$), to assure that we do not overproduce highest
mass, and therefore the easiest to detect, BH-BH binaries. As it turns out,
the change of IMF slope does not significantly alter the detection rate.
Farther steepening of the IMF ($\alpha=-3$) decreases the expected rate by
half, while flatter IMF assumption ($\alpha=-1$) increases the expected rate
by a factor of two. Note that we maintain a flat IMF between $1\,{\rm M}_\odot$
and $100\,{\rm M}_\odot$ when calculating $f_{sim}$ -- the fraction of stars
that we simulate out of the total population. The observed rate scales linearly
with $f_{sim}$. We also fixed the maximum mass of the Population III stars to
$500\,{\rm M}_\odot$. Had we allowed for possibility of star formation with
the higher mass the rate would increase.

The estimate of the lifetime of the black hole binaries can be strongly
affected if the binaries are hardened by interactions in dense stellar
environments. Our simulations show that a large number of black hole binaries
should be formed with the lifetimes in excess of the Hubble time.
Interactions in dense systems may significantly shorten their lifetimes.
We demonstrated above that hardening does not affect
strongly the expected rate. Another possibility arises that the interactions
disrupt some of BH-BH binaries. This would deplete the population of wide
systems. None of the two effects, unless operating on extremely short
timescales, can affect much systems with shortest coalescence timescales.
Therefore, a combination of the two effects may constrain the mergers of
Population III BH-BH binaries to large redshifts.

In order to estimate the $S/N$ for advanced LIGO we have used the formulae
of \citet{1998PhRvD..57.4535F}. The predicted values of $S/N$ may still
change when more realistic noise curves and binary gravitational wave form
templates are known. Because of large masses of the binaries considered in
this work the changes in the low-frequency range are most important.
The typical ringdown frequency is $\nu_{qnr} \approx 90 (300{\rm M}_\odot /M)
$\,Hz, so the rate is most sensitive to the detector performance in the low
frequency region. The scaling of the rate with the change of $S/N$
normalization can be read off Figure~\ref{rsn}.

The influence of the evolutionary model assumptions on detection
rate was tested.
We changed the CE efficiency (increase/decrease by factor of 2); altered
evolution through stable RLOF phases from fully conservative to
non-conservative cases; changed the specific angular momentum of the matter
leaving the system during RLOF (increase/decrease by factor of 2) and
varied the critical mass ratio over which the dynamical instability
develops (from $q_{\rm crit}=2$ in standard model to $q_{\rm crit}=1-3$).
The detection rate was decreased at most by a factor of 3 in the above
models. Therefore, the predicted rate does not depend strongly on the
binary evolution within the model assumptions.

Summarizing, in our calculations we have tried to use rather conservative
assumptions in order not to overestimate the detection rate of Population
III BH-BH mergers. If several of the assumptions and values of the
model parameters are changed within reasonable limits, we  still obtain
a significant detection rate. Even if the rate predicted for our already
conservative model is decreased by 2-3 orders of magnitude to allow for
different aforementioned uncertainties, we still are left with several
strong events per year for advanced LIGO detector.
For initial LIGO phase the detection rate falls below one event per year.

We have shown that Population III stars may lead to formation of a large
number of binaries containing intermediate-mass black holes.
A significant fraction of such systems has coalescence times smaller than
the Hubble time. Coalescences of intermediate-mass BH-BH binaries should be
detectable by advanced interferometric gravitational wave detectors during
the ringdown phase, and possibly also merger phase, provided that accurate
templates are available. Given the large expected rate of observed
coalescences such events could be the primary targets for LIGO burst 
\vspace*{-0.35cm} search.

\acknowledgements
We thank S.Hughes, E.Flanagan, R.Taam and A.G\"urkan
for useful comments and Northwestern University Theory Group for 
hospitality (TB). We acknowledge support of the KBN grant 
PBZ-KBN-054/P03/2001.

%\bibliographystyle{aa}
%\bibliography{pop3}

\end{document}